\documentclass[letterpaper,prb,twocolumn,showpacs,preprintnumbers,amsmath,ammsymb,superscriptaddress,footinbib,citeautoscript,citeautoscript]{revtex4}
\usepackage[bookmarks=false]{hyperref}
\usepackage{amsmath}
\usepackage{amssymb}
\usepackage{bm}
\usepackage{amsthm}
\usepackage{amsfonts}
\usepackage{latexsym}
\usepackage{graphicx}
\usepackage{microtype}
\usepackage{dcolumn}

\begin{document}

\title{Lattice normal modes and electronic properties of the correlated
metal LaNiO$_3$}

\author{Gaoyang Gou}\email{gaoyang@sas.upenn.edu}
\affiliation{%
The Makineni Theoretical Laboratories, Department of Chemistry,
University of Pennsylvania, Philadelphia, PA 19104-6323, USA}
\author{Ilya Grinberg}\email{ilya2@sas.upenn.edu  }
\affiliation{%
The Makineni Theoretical Laboratories, Department of Chemistry,
University of Pennsylvania, Philadelphia, PA 19104-6323, USA}
\author{Andrew M.\ Rappe}\email{rappe@sas.upenn.edu}
\affiliation{%
The Makineni Theoretical Laboratories, Department of Chemistry,
University of Pennsylvania, Philadelphia, PA 19104-6323, USA}
\author{James M.\ Rondinelli}\email{jrondinelli@coe.drexel.edu}
\affiliation{Department of Materials Science \&
Engineering, Drexel University, Philadelphia, Pennsylvania 19104,
USA}
\affiliation{%
X-Ray Science Division, Argonne National Laboratory, Argonne, IL
60439, USA}

\date{\today}

\begin{abstract}\sloppy
We use density functional theory (DFT) calculations to
study the lattice vibrations and electronic properties of
the correlated metal LaNiO$_3$.
To characterize the rhombohedral to cubic structural phase transition of perovskite
LaNiO$_3$, we examine the evolution of the Raman-active phonon modes
with temperature.
We find that the $A_{1g}$ Raman mode, whose frequency is sensitive
to the electronic band structure, is a useful signature to
characterize the octahedral rotations in rhombohedral LaNiO$_3$.
We also study the importance of electron--electron correlation
effects on the electronic structure with two approaches which go
beyond the conventional band theory (local spin density
approximation): the local spin density$+$Hubbard $U$ method
(LSDA$+U$) and hybrid exchange-correlation density functionals which
include portions of exact Fock-exchange.
We find the conventional LSDA accurately reproduces the delocalized
nature of the valence states in LaNiO$_3$ and gives the best
agreement to the available experimental data on the electronic
structure of LaNiO$_3$.
Based on our calculations, we show that the electronic screening
effect from the delocalized Ni 3$d$ and O-2$p$ states mitigate
the electronic correlations of the $d^7$ Ni cations, making
LaNiO$_3$ a weakly correlated metal.
\end{abstract}

\maketitle
\sloppy

\section{Introduction}
Conducting electrode materials are critical elements in the design
of ultra-thin ferroelectric
devices,\cite{Stengel/Vanderbilt/Spaldin:2009,Sai/Kolpak/Rappe:2005,Kolpak/Sai/Rappe:2006,Al-Saidi10p155304}
magnetoresistive elements\cite{Herranz/Fontcuberta_et_al:2004}, and
magnetoelectric multiferroic
memories.\cite{Garcia/Barthelemy_et_al:2010}
The performance of perovskite-based heterostructures is intimately
related to the electronic and atomic coherency across the
electrode--oxide interface.\cite{Mathews/Ramesh:1997}
For this reason, suitable metallic perovskite oxides that are structurally
compatible with the active functional layers are highly desirable.
LaNiO$_3$ (LNO) has found widespread use in this capacity
\cite{Kim_et_al:2007,Jain/Kang/Jia:2006,Murugavel/Rao:2000}, because
it is a 3$d$ transition metal oxide that shows no metal-insulator or
spin state transitions.\cite{Lacorre/Torrance:1992, Medarde:1997}
At present, it is also actively being pursued as the
\textit{functional} oxide component in devices that could harness an
electric-field tunable Mott insulator transition (so called
``Mottronic"
applications\cite{Scherwitzl/Triscone_etal:2009,Son/Stemmer_etal:2010,Scherwitzl/Triscone_etal:2010}),
because it is the end-member of the rare-earth nickelate
series---compounds with small charge-transfer gaps that are highly
susceptible to temperature\cite{Torrance/Niedermayer_etal:1992} and
pressure-induced\cite{Canfield_et_al:1993,Obradors/Torrance_et_al:1993}
electronic phase transitions. %%

Metallicity and magnetism in 3$d$ transition metal (TM) oxides are
strongly dependent on the valence bandwidth, which originates from
the hybridization between the TM 3$d$ and O 2$p$ orbitals.
\cite{Lacorre/Torrance:1992}
In perovskite oxides like LNO, the hybridization derives from
the structural connectivity of the NiO$_6$ octahedral units throughout the crystal.
Small changes in the Ni--O--Ni bond angles and the Ni--O bond lengths, therefore,
can dramatically alter the electronic properties.
For example, a tunable insulator-metal (IM) transition is obtained
from the isovalent substitution\cite{Catalan:2008} of La with rare-earth
elements: The charge-transfer gap closes and the IM-transition becomes accessible
above room temperature as the rare-earth ionic radius increases
and straightens the Ni--O--Ni bond angle.%%

In addition to changes in the structural degrees of freedom, the
small spatial extent of the 3$d$ orbitals also reduce the valence
bandwidth.
This sufficiently enhances electron--electron correlation effects such
that conventional band metals are often rendered insulating.\cite{Imada/Fujimori/Tokura:1998}
In LaNiO$_3$, however, the strong Ni 3$d$ -- O 2$p$ covalent
interactions---formally it contains Ni$^{3+}$ cations in a low-spin
3$d^7$ configuration ($t_{2g}^6e_g^1$)---are anticipated to reduce
the correlation effects.\cite{Goodenough/Raccah:1965}
Nonetheless, there are clear signatures indicative of
important electron--electron interactions in the
$T^2$ dependence of the resistivity and heat capacity measurements.\cite{Rajeev_et_al:1991,Lacorre/Torrance:1992}
Magnetic susceptibility data also reveal enhanced Pauli
paramagnetism and effective carrier
masses.\cite{Vasanthacharya_et_al:1984,Sreedhar/Spalek_et_al:1992,Torrance/Niedermayer_etal:1992,Xu_et_al:1993,Zhou/Goodenough_et_al:2000}
Consistent with those studies, temperature dependent x-ray photoemission finds
that the spectral weight of the Ni $e_g$ band at the Fermi level increases upon
cooling.\cite{Horiba_et_al:2007}
A fundamental question regarding the intrinsic properties of
LaNiO$_3$ therefore still remains: which factors of the interwoven
(correlated) electronic and atomic structure support its metallic state? %%

In this work, we perform first-principles density-functional theory
(DFT) calculations to investigate how hybridization between the Ni
3$d$ and O 2$p$ states and the structurally correlated NiO$_6$
octahedral framework respond to electron-electron interactions.
We first examine the temperature-dependent rhombohedral to cubic structural
phase transition using a Landau formalism\cite{Toledano:1987} and {\it ab initio}-derived
phenomenological coefficients obtained from the conventional
band theory [local spin density approximation, (LSDA)].
In the rhombohedral phase, we compute the evolution of the
Raman-active phonon modes with temperature and find the LSDA results
give good agreement with experiment.
We then show how the $A_{1g}$ Raman mode, which describes the
rotation of adjacent NiO$_6$ octahedra, can be used as a structural
indicator for the low-temperature rhombohedral phase. %%

To study the coupling between the lattice modes and the electronic
structure, we compare conventional LSDA results with three other
approaches designed to improve accuracy: the local spin
density$+$Hubbard $U$ (LSDA$+U$) method, and two hybrid
exchange-correlation functionals, PBE0
\cite{Perdew_et_al:1996,Wu09p085102} and
HSE\cite{Heyd/Scuseria/Ernzerhof:2003,Heyd/Scuseria/Ernzerhof:2006},
which contain mixtures of Fock-exchange added to the generalized
gradient corrected DFT-functional of Perdew, Burke and Ernzerhof
(PBE).\cite{Perdew/Burke/Ernzerhof:1996}
We then examine the electronic structure of LaNiO$_{3}$ by comparing
the results obtained from the various functionals with the
experimental photoemission spectroscopy (PES) and x-ray
photoelectron spectroscopy (XPS) data.
We find that the screening effects
originating from the hybridized O~2$p$ and Ni~3$d$ electrons
are sufficiently strong that they reduce the electronic
correlations in LaNiO$_3$, making it a weakly correlated metal. %%
\begin{figure}[b]
\centering
\includegraphics[width=0.45\textwidth]{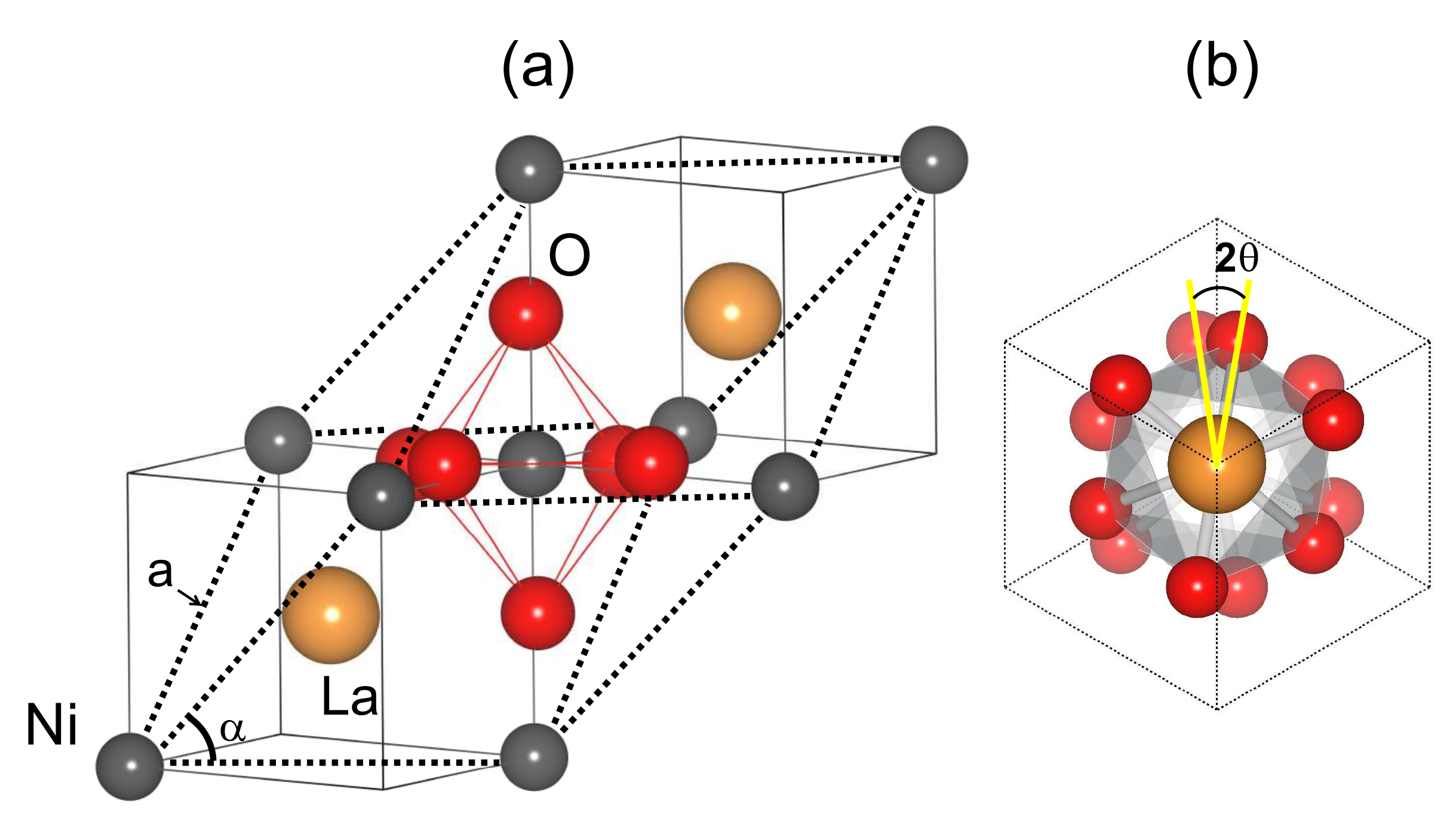}
\caption{(Color)
(a) The crystal structure of rhombohedral $R\bar{3}c$ LaNiO$_3$
possesses anti-phase rotations ($a^-a^-a^-$ tilt system)
of adjacent NiO$_6$ octahedra with angle $\theta$ (b)
about the [111]-trigonal axis.
The relationship between the pseudo-cubic perovskite cell (solid lines) and
the rhombohedral lattice vectors (dashed lines) of length $a$ is also illustrated.
}
\label{Fig.1}
\end{figure}
\section{Crystal Structure and Raman Modes}

The crystallographic tolerance factor \cite{Goldschmidt:1926} for
LaNiO$_3$ is $t$=0.97 and indicates that the aristotype cubic
phase is susceptible to octahedral rotations because of an
under-bonded La--O coordination.\cite{Sreedhar/Spalek_et_al:1992}
At ambient conditions, bulk LaNiO$_{3}$ crystallizes in a rhombohedral
structure with space group $R\bar{3}c$ [Fig.\ \ref{Fig.1}(a)]
and is related to the cubic perovskite through a trigonal lattice
distortion along the [111]-body diagonal
that doubles the primitive unit cell\cite{Lacorre/Torrance:1992}.
It also exhibits octahedral rotations, which are equal in magnitude
and alternate in ``sense'' about each Cartesian direction; this tilt
system is classified according to Glazer
notation\cite{Glazer:1972,Stokes/Kisi_et_al:2002} as $a^-a^-a^-$
[Fig.\ \ref{Fig.1} (b)].
Similar to other rhombohedrally distorted perovskites,
\cite{Abrashev:1999} LaNiO$_{3}$ undergoes a temperature-induced
rhombohedral-to-cubic phase transition\cite{Angel/Zhao/Ross:2005}
upon heating: the NiO$_6$ octahedral rotation angle $\theta$ reduces
continuously to zero in the cubic phase. %%

\begin{figure}[t]\vspace{-4pt}
\centering
\includegraphics[width=0.49555\textwidth]{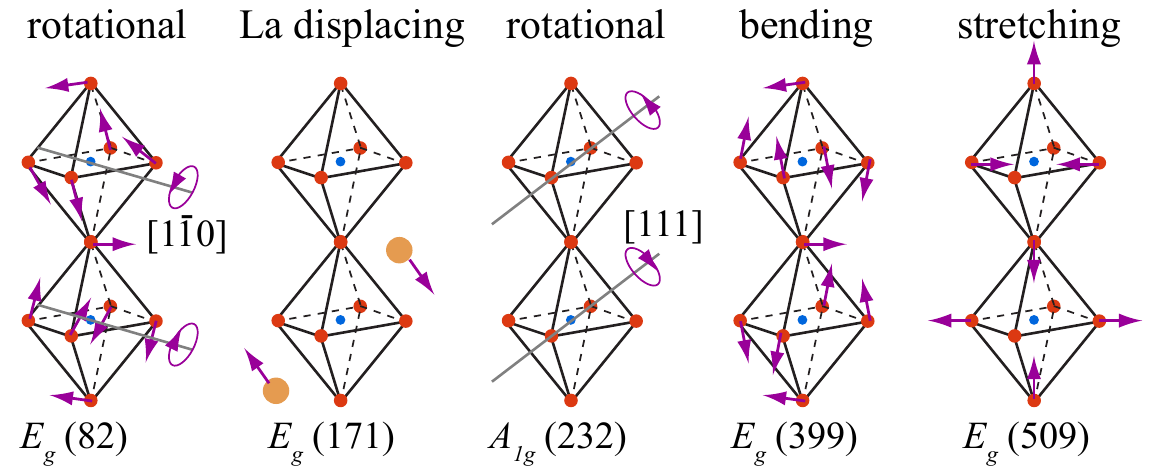}
\caption{(Color)
Illustration of the displacement patterns for the Raman-active lattice normal
modes in the $R\bar{3}c$ structure. The corresponding symmetry labels and our
calculated LSDA frequencies (in cm$^{-1}$) are given for reference.}
\label{fig:raman_modes}
\end{figure}

In the $R\bar{3}c$ space group, the La cations occupy the $2a\,
(\frac{1}{4},\frac{1}{4},\frac{1}{4})$ Wyckoff positions while the
Ni cations occupy the $2b\, (0,0,0)$ positions.
The oxygen atoms are at the $6e\, (x,\bar{x}+\frac{1}{2},\frac{1}{4})$ site, where
$x$ is the only free internal structural parameter that sets the rotation
angle of the NiO$_6$ octahedra.
These 10 atoms  in the primitive rhombohedral unit cell give rise to
30 zone-center vibrational modes with the irreducible
representations (irreps):
\[
\Gamma = A_{1g}+3A_{2g}+4E_{g} +2A_{1u}+4A_{2u}+6E_{u} \,.
\]
where the $E_g$ and $E_u$ modes are two-dimensional irreps.  The
infrared-active and acoustic modes transform as the $4A_{2u} +
6E_u$, while the Raman active modes are given as $A_{1g} + 4E_g$.
The $A_{1g}$ and $E_g$ Raman modes of the rhombohedral structure are mainly related to
collective modes of the oxygen octahedral network (Figure \ref{fig:raman_modes}):
The $A_{1g}$ mode describes the rotations of the NiO$_6$ about the trigonal [111]-axis,
and the  $E_g$ modes describe anti-parallel La displacements, Ni--O bond bending,
stretching, and octahedral rotations about axes perpendicular to the [111]-direction.

\section{Theoretical Methods}

\subsection{LSDA}

We use two DFT implementations in this study, the {\sc quantum
ESPRESSO} code ({\sc QE}) \cite{QuantumEsprsso:2009} and the Vienna
{\it ab initio} Simulation Package ({\sc vasp})
\cite{Kresse/Furthmuller:1996a,Kresse/Joubert:1999}.
Our reference electronic structure model to which we compare our
advanced treatments of correlation effects is the frequently used
local-spin density approximation (LSDA) exchange-correlation (XC)
functional.\cite{Perdew/Zunger:1981} Both DFT codes use the
Perdew-Zunger parametrization\cite{Perdew/Zunger:1981} of the
Ceperley-Alder data \cite{Ceperley/Alder:1980} for the
XC-functional.
In all calculations, we constrain a collinear spin configuration for the
Ni ions.
The atomic positions in the rhombohedral structure are optimized by
starting from the positions reported in Ref.\
\onlinecite{Lacorre/Torrance:1992}, and the ionic coordinates are
relaxed until the Hellmann-Feynman forces on the atoms are less than
0.1 meV~\AA$^{-1}$. %%

{\it {\sc vasp} Details.$\quad$}%
We treat the core and valence electrons for all calculations in {\sc vasp}
using the projector augmented wave (PAW) method \cite{Blochl:1994}
with the following valence electron configurations:
$5p^65d^16s^2$ (La), $3p^63d^94s^1$ (Ni), and $2s^22p^4$ (O) and a
650~eV plane wave cutoff.
We use a Gaussian smearing of 0.05~eV over a $7\times7\times7$
Monkhorst-Pack $k$-point mesh \cite{Monkhorst/Pack:1976} centered at
$\Gamma$ for the Brillouin zone (BZ) integrations (172 points are
sampled in the irreducible BZ) in the 10-atom rhombohedral unit
cell. %%

We obtain the phonon frequencies at the $\Gamma$-point
by calculating total energies with respect to atomic displacements
from the reference $R\bar{3}c$ structure.
In this frozen-phonon method a series of small (symmetry
inequivalent) atomic displacements are imposed along different
Cartesian directions.
We calculate the dynamical matrix from the Hellmann-Feynman
forces\cite{Gonze/Lee:1997} induced on the ions after making the
small positive and negative displacements (to remove any quadratic
effects) about the reference structure positions from total energy
DFT calculation.
Diagonalization of the dynamical matrix yields the atomic displacement patterns
(eigenvectors) and phonon mode frequencies (eigenvalues).
{\it QE Details.$\quad$}%
In our plane-wave calculations with the {\sc QE} code, we use
norm-conserving\cite{Rappe:1990} optimized designed nonlocal
\cite{Ramer:1999} pseudopotentials.
The following valence electron configurations are used:
$5s^25p^64f^05d^06s^06p^0$(La), $3d^94s^14p^0$ (Ni) and $2s^22p^4$
(O).
Note, In both {\sc vasp} and {\sc QE}, partial core corrections (PCC)
are included in the Ni pseudopotentials following the
prescription in Ref.~\onlinecite{Louie82p1738}

We sample the BZ using an $8\times8\times8$ Monkhorst-Pack $k$-point
mesh. A plane-wave energy cutoff of 50 Ry was used for calculation.
Good convergence can be reached, as test calculations with higher
energy cutoff (up to 75 Ry) yield the same results.
We obtain phonon modes and displacement vectors using density
functional perturbation theory
(DFPT),\cite{Baroni/deGironcoli/DalCorso:2001,Gonze:1995} by
constructing the interatomic force constants from the dynamical
matrices obtained on a uniform $4\times4\times4$ $q$-point grid.
The phonon frequencies and vibrational eigenvectors at arbitrary $q$
vectors are then calculated from diagonalization of the dynamical
matrix. %%

\begin{table}[b]
\caption{%
Structural parameters obtained for $R\bar{3}c$ LaNiO$_3$ within the
LSDA from the two codes used in this work and their comparison with
experimental data (cf.\ Ref.\ \onlinecite{Lacorre/Torrance:1992}).
In the rhombohedral setting, the rotation angle of adjacent NiO$_6$
octahedra along the trigonal axis is given as
$\theta=\textrm{arctan}(2\sqrt{3}u$), where $x=u+\frac{3}{4}$.}
\begin{ruledtabular}
\centering
\begin{tabular}{lddd}%
                    &  & \multicolumn{2}{c}{Theory} \\
\cline{3-4}
 & \multicolumn{1}{c}{Experiment} &  \multicolumn{1}{c}{\sc vasp}   &   \multicolumn{1}{c}{QE}    \\
\hline
$x$                                 & 0.7968     & 0.787  & 0.801  \\
$a$ ({\AA})                         & 5.3837     & 5.303  & 5.324  \\
$\alpha_{\textrm{rho}}$ ($^\circ$)  & 60.8596    & 60.72  & 61.39  \\
$d$(Ni--O) ({\AA})                  & 1.933      & 1.896  & 1.923  \\
$\angle$Ni--O--Ni ($^\circ$)        & 164.8      & 167.9  & 163.3  \\
$\theta$ NiO$_6$ rotation ($^\circ$)& 9.2        & 7.35   & 10.08  \\
$\Omega$~({\AA$^3$}/f.u.)           & 56.2386    & 53.57  & 55.03  \\
\end{tabular}
\end{ruledtabular}
\label{tab:rho_structures}
\end{table}

We point out here that both implementations of DFT at the LSDA level
give good structural agreement (Table \ref{tab:rho_structures}).
This fact is important for the comparison study we make later which uses
various functionals available in the different codes.
Each code yields a small  underestimation of the experimental
lattice constant which is typical for the LSDA.
The {\sc QE} code more closely reproduces the experimentally
measured Ni--O--Ni rotation angles, whereas the {\sc vasp} code
gives a better rhombohedral angle ($\alpha_{\textrm{rho}}$).
For the electronic and magnetic structure of LNO, both codes predict
LNO as a non-magnetic metal, which is the reasonable representation
of the experimentally observed Pauli paramagnetic
ground-state.\cite{Sreedhar/Spalek_et_al:1992} Detailed electronic
properties for LNO will be presented in Sec.~\ref{sec:results}.
Here we note that PCC for Ni is crucial to obtain the correct
ground-state of paramagnetic LNO. Without PCC, LSDA calculations
yield a spin-polarized solution with a non-zero local magnetic
moment on the Ni cations.

\subsection{LSDA$+$$U$}
A complete description of correlated 3$d$ transition metal oxides with
narrow valence bandwidths is challenging within a density functional
approach due to the limitations of standard exchange-correlation
potentials in describing localized electronic states.
To remedy this problem, we use ``beyond-LSDA'' techniques, starting
with the LSDA+$U$ approach.
This method is generally regarded to be the most computationally
feasible means to reproduce the correct ground-states in correlated
systems \cite{Anisimov/Lichtenstein:1997}.
In this formalism, the LSDA energy functional is expanded to include
an additional  on-site orbital-dependent energy term cast as a
Hubbard repulsion $U$ and an intratomic Hund's exchange energy $J$. %%
To reduce the ambiguity in the definition of the LSDA+$U$ parameters,
we use the spherically averaged
form of the rotationally invariant LSDA plus Hubbard $U$
method \cite{Anisimov/Lichtenstein:1997} introduced
by Dudarev {\it et al.},\cite{Dudarev/Sutton_et_al:1998} with
only one effective Hubbard term, $U_{\rm eff}=U-J$.
We treat the double-counting term
within the fully localized limit \cite{Sawatzky:1994}.

The change in total energy $E_{U}$ for including the Hubbard
correction to the exchange-correlation potential is
\begin{equation}
E_{U} (J=0) = \frac{U_{\rm eff}}{2}\sum_{i}\sum_{m \sigma}{n^{i}_{m
\sigma}(1-n^{i}_{m \sigma})}\, ,
\end{equation}
where $n^{i}_{m \sigma}$ are the spin ($\sigma$) and orbital ($m$)
occupation numbers at site $i$.
In the limit where the occupation matrices are integer and diagonal,
the LSDA$+U$ correction can be understood as a shift in the occupied
orbitals ($n_m=1$) by $-U/2$ to lower energy and
by $+U/2$ higher in energy  for unoccupied orbitals ($n_m=0$).
In this study we examine $U_{\rm eff}$ values of 0, 3 and 6~eV for
the  correlated Ni 3$d$ orbitals states; the standard LSDA corresponds to a
$U_{\rm eff}=0$ eV.
We note that throughout the remainder of the manuscript, $U$ denotes
the effective Hubbard parameter.
We will discuss the structural changes induced by varying $U$ in the
subsequent sections. %%

\subsubsection{Self-consistent Hubbard $U$}
We also calculate a self-consistent $U$ parameter for LaNiO$_3$
following the scheme developed by Cococcioni and de Gironcoli
\cite{Cococcioni/Gironcoli:2005}.
Their approach relies on the fact that the potential $U$ restores
the correct piece-wise linear behavior of the system's total energy
as function of electron number, whereas the LSDA functional
incorrectly predicts parabolic dependence of energy on occupation
number.
The effective interaction parameter $U$ is deduced from
Janak's theorem and linear response theory as
\begin{equation}
U=(\chi_{0}^{-1}-\chi^{-1})_{ii}\, , \textrm{with}\,  \chi_{ij}=\frac{dn^{i}}{d\alpha_{j}}\, .
\end{equation}
Here $n^{i}$ is the occupation number of the localized levels at
site $i$ and $\alpha_{j}$ represents the potential shift applied on
the localized orbital at site $j$. The response matrix $\chi_0$
describes the noninteracting contribution to the band structure
after application of a potential shift $\alpha$, while $\chi$
represents the fully self-consistent response to the same potential
shift.
In practice, the first term $\chi_0$ is computed from the first iteration in the
self-consistent (SCF) electronic minimization.
We compute the linear response of the occupation number $n^{i}$
using the LSDA functional and norm-conserving pseudopotentials
within the {\sc QE} package.
The full response matrices $\chi_0$ and $\chi$ are then computed by
performing the linear response calculation within a $2\times 2\times
2$ pseudo-cubic supercell (40-atom cell, containing 8 Ni
atoms), which is large enough to give a converged $U$ value and avoid
spurious interactions from the local potential $\alpha$ on neighboring
Ni sites.

\begin{figure}
\centering
\includegraphics[width=0.48\textwidth]{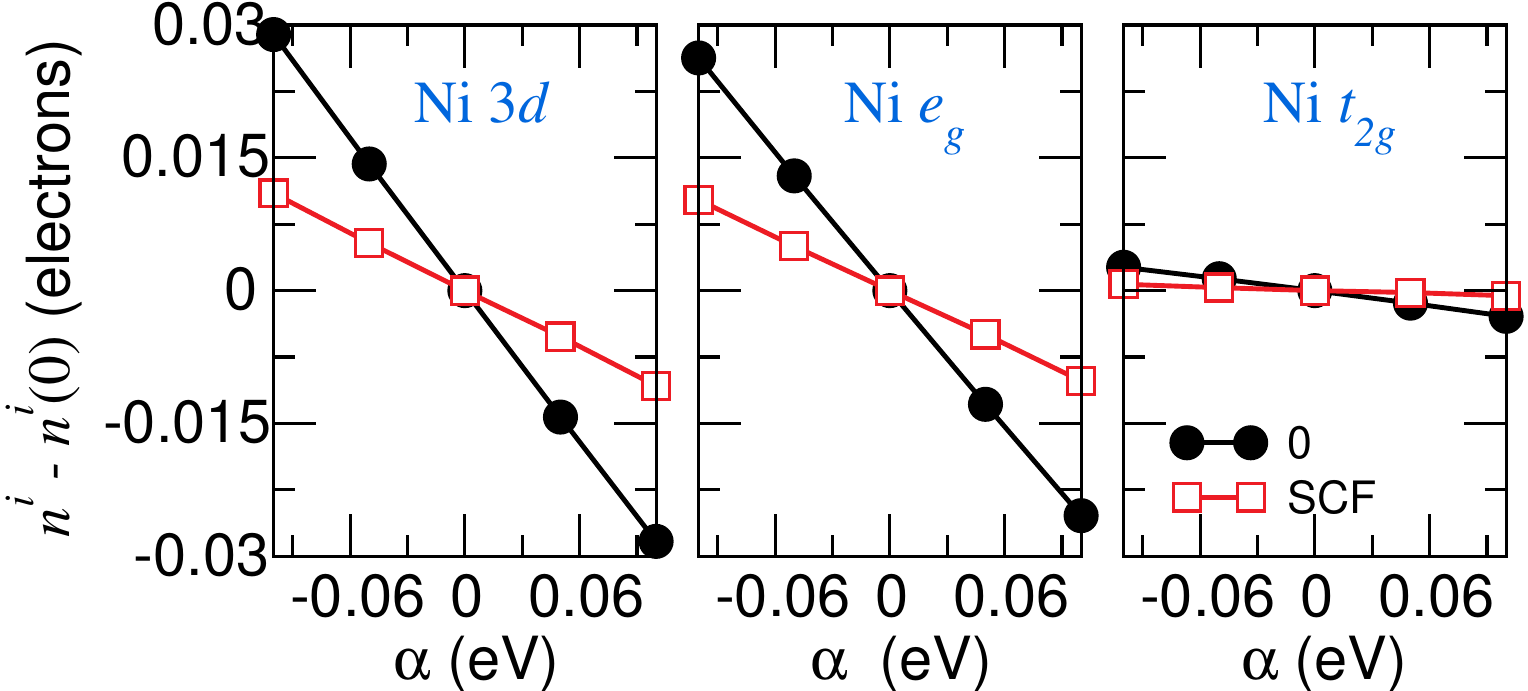}
\caption{(Color) The occupation numbers n$^{i}$ for the total Ni 3$d$,
$e_{g}$ and $t_{2g}$ orbitals in LaNiO$_{3}$, with respect to potential shift
$\alpha$ for the initial (0)~(filled circles) and self-consistent~(SCF) (open squares)
electronic configurations.
The slopes of the linear data fits are used to construct the
$\chi_0$ and $\chi$ response matrices.
The occupation numbers are normalized by substracting off the
occupation values $n_{i}(0)$ with a zero potential shift ($\alpha=0$).} \label{Fig.2}
\end{figure}
The low-spin Ni 3$d$ orbital occupation anisotropy---a fully filled
$t_{2g}$ manifold and a quarter-filled e$_{g}$ shell---suggests that
the effective Coulomb repulsion experienced by an electron on
different orbitals should also be substantially
different.\cite{Pickett_et_al:1998}
For this reason, we further decompose the orbital occupancy $n^i
\rightarrow n^i_{m}$ ($m$ is the angular momentum quantum number,
and orbital occupation $n^i_{m}$ here refers to L\"owdin atomic
charge) into crystal-field split $d$-manifold contributions by
projecting the valence states $|\Psi^{\sigma}_{k\nu}\rangle$ onto
atom-centered sites $i$ with a L\"owdin orthogonal atomic basis set
$|\phi^{i}_{m}\rangle$ as
\begin{equation}
V_{U}|\Psi^{\sigma}_{k\nu}\rangle = %\frac{\delta
\frac{U}{2}\sum_{i}\sum_{m
\sigma}{(1-2n^{i}_{m
\sigma})|\phi^{i}_{m}\rangle\langle\phi^{i}_{m}|
\Psi^{\sigma}_{k\nu}}\rangle\, .
\end{equation}
By replacing the orbital occupancy $n^i$ in Eq. (2) with $n^i_{m}$,
we can calculate the effective Hubbard term $U$ for the two
independent Ni-$d$ manifolds: $U(e_{g})$ and $U(t_{2g})$
respectively.
Since the $t_{2g}$ manifold is fully occupied [$n(t_{2g})\approx1$],
the linear response calculation of $U(t_{2g})$ requires that the
contribution from these states to the total Coulomb interaction be
nearly zero.
Within the numerical noise of our calculations we find this to be
the case.
Fig.\ 3 shows that difference in the $\chi_0$ and $\chi$
response functions (slopes) of the $t_{2g}$ states is practically negligible.
The Ni $e_{g}$ states, however, which are partially filled
have a substantial non-zero interaction, $U(e_{g})$.
These states lead to a self-consistent Hubbard $U$ value of 5.74 eV
for LaNiO$_3$.
Recently Nohara {\em et al.} studied LaNiO$_{3}$ with LSDA$+U$ and
$GW$ methods,\cite{Nohara:2009} and fitted their calculated energy
spectra to experimental XPS and x-ray adsorption spectroscopy (XAS)
data\cite{Abbate:2002} to obtain a Coulomb interaction of $U=7.0$~eV
and an exchange interaction $J=1.3$~eV for the Ni $d$-manifold.
These values correspond to an effective Hubbard $U$ term ($U_{\rm
eff}=5.7$~eV) which is consistent with our linear-response
calculation of $U$. %%

\subsection{Hybrid Functionals}
Recent studies
\cite{Perdew/Burke/Ernzerhof:1996,Wu/Rappe/Car_et_al:2009,Heyd/Scuseria/Ernzerhof:2003,Heyd/Scuseria/Ernzerhof:2006}
have demonstrated that local or gradient-corrected density
functionals for exchange do not closely reproduce Hartree-Fock
calculations, and that inclusion of some exact Fock ($\rm F$)
exchange improves electronic structure properties including band
gaps, orbital  localization and electronic polarizations.\cite{Stroppa/Picozzi:2010}
Accordingly, in the present work, we
use the PBE0 and HSE hybrid functionals as implemented in the {\sc
QE} package and then contrast them to the LSDA.
The exchange terms ($V_x$) in the PBE0 functional are constructed by
mixing 25\% of exact-exchange with 75\% GGA-PBE exchange
\cite{Perdew_et_al:1996}, while the electron correlation ($V_c$) part
is represented using only the correlation components from the PBE functional
\cite{Perdew/Burke/Ernzerhof:1996}:
\begin{equation}
V_{xc}^{\rm PBE0}=\frac{1}{4} V_{x}+\frac{3}{4}V_{x}^{\rm
PBE}+V_{c}^{\rm PBE}\, .
\end{equation}
The one-quarter mixing parameter of exact exchange in the PBE0
functional was obtained by fitting to the atomization energies of a
large database of molecules.\cite{Perdew_et_al:1996} For periodic
bulk systems, however, it is argued that the best fraction of exact
exchange for modeling solid-state electronic structure is highly
system-dependent.\cite{Perdew_et_al:1996,Ernzerhof:1996}

The HSE hybrid functional
\cite{Heyd/Scuseria/Ernzerhof:2003,Heyd/Scuseria/Ernzerhof:2006} is
less computationally demanding than PBE0 (unless the recent Wannier
function implementation described in Ref.~\onlinecite{Wu09p085102}
is used), since it avoids the slowly decaying Fock-exchange
interactions by substituting  part of the long-range Coulomb-kernel
with a density functional analog.
This approximation makes the HSE functional
more efficient for plane-wave calculations of periodic systems.
The expression for the exchange-correlation potential is given as
\begin{equation}
V_{xc}^{\rm HSE}=\frac{1}{4} V^{\rm sr,\mu}_{x}+\frac{3}{4}V_{x}^{\rm
PBE,sr,\mu}+ V_{x}^{\rm PBE,lr,\mu}+V_{c}^{\rm PBE}\, ,
\end{equation}
where $\mu$ is the parameter that controls the decomposition of the
Coulomb kernel into short-range (sr) and long-range (lr) exchange
contributions.
This type of screened exact-exchange is absent in the PBE0 functional.
In this work, we follow the HSE03 parametrization\cite{Heyd/Scuseria/Ernzerhof:2003} which
sets the cut-off distance to $\mu=0.109$~\AA$^{-1}$.
Similar to PBE0, the Fock-mixing parameter in HSE varies
approximately in proportion to the inverse static dielectric
constant ($\varepsilon_{\infty}^{-1}$), and therefore also becomes highly
system dependent.
For metals with excellent screening, $\varepsilon_{\infty}^{-1}$
approaches zero.
Therefore PBE0 or HSE calculations which incorporate some
Fock-exchange may lead to severe overestimation in orbital
bandwidths and spin-exchange splitting parameters for such
materials.\cite{Paier:2006} %%

In our hybrid functional calculations on metallic LaNiO$_{3}$, we
use a dense Brillouin zone sampling ($8\times8\times8$ $k$-point
grid) for accurate evaluation of the Fock-exchange operator in
reciprocal space. \cite{Paier:2006} Due to the heavy computational
cost, we are unable to perform the structural optimization for LNO
using the hybrid functionals in the present work.
According to our test calculations using the LSDA, PBE and LSDA$+U$
functionals, the subtle differences between the ground-state atomic
positions produce negligible changes to the computed electronic band
structure. We therefore choose the LSDA optimized atomic structure
as the ground-state structure for use in the hybrid functional
calculations.

\section{Results and Discussion\label{sec:results}}
In this section, we present our first-principles results on the
structural, vibrational and electronic properties of bulk
LaNiO$_{3}$ obtained using the various XC-functionals.
This section is organized as follows: we first investigate the
structural phase transition and temperature-dependent Raman
phonon modes of LaNiO$_3$ using the conventional LSDA approximation.
We then examine how the atomic structure, Raman phonon modes and
electronic properties of LaNiO$_3$ are modified with various
treatments of electron-electron correlation effects. Finally, by
comparing the densities-of-states results obtained from the
different functionals to the experimental spectroscopic data, we
identify which XC-functional best reproduces the intrinsic
electronic properties of LaNiO$_{3}$.

\subsection{Structural Phase Transition}
The temperature-induced rhombohedral-to-cubic phase transition in
LaNiO$_{3}$ manifests as a cooperative rotation of NiO$_6$ octahedra
along the trigonal lattice axis.\cite{Chaban/Kreisel_et_al:2010}
In this section, we show how the structural transition to the
rhombohedrally distorted perovskite phase is characterized by a
specific soft mode which has a frequency that collapses to zero as the
system approaches the cubic phase. \cite{Abrashev:1999}
We note that all results in this section are obtained with the LSDA
XC-functional and NC-pseudopotentials as implemented in {\sc QE}. %%

\subsubsection{Minimal Landau Model\label{sec:landau}}
\begin{figure}
\centering
\includegraphics[width=0.49\textwidth]{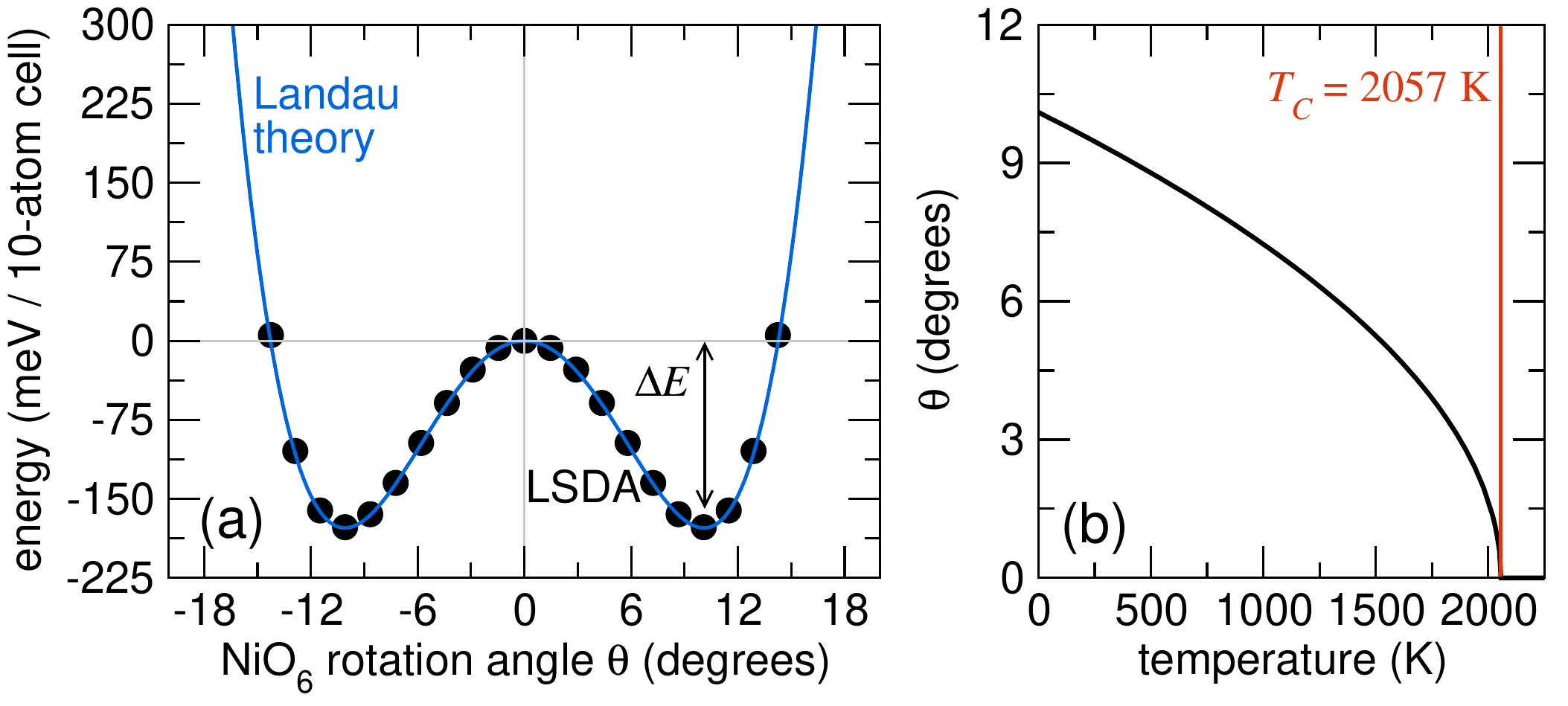}
\caption{(Color) (a) Landau free energy $G$($\theta,T=0$~K) of LaNiO$_{3}$ as
a function of the order parameter $\theta$. Solid lines are
calculated using Landau theory and the filled symbols correspond to
the LSDA total energy results. (b) Equilibrium order parameter
$\theta$ as a function of temperature; the second order phase transition
occurs at $T=T_{C}$.} \label{Fig.3}
\end{figure}

In the Landau theory framework for phase transitions, the free
energy of the system is expanded in powers of an order parameter
that characterizes the transition.
From our previous discussion of the low-temperature $R\bar{3}c$
crystal structure of LaNiO$_{3}$, the only free \textit{internal}
parameter is the oxygen position $x$ at the $6e$ Wyckoff site.
Therefore, the natural order parameter to characterize the
rhombohedral-to-cubic phase transition is a structure-adapted form
of the free Wyckoff position: the NiO$_6$ octahedra rotation angle
$\theta$ (Fig.\ \ref{Fig.1}).
We  expand the free energy $G$ in even powers (up to fourth
order) of this rotation angle order parameter $\theta$ as:
\begin{equation}
G(\theta, T)=G_{0}(T)+\kappa(T-T_{C})\theta^{2}+\lambda\theta^{4}
\label{eq:landau}
\end{equation}
where $\kappa$ and $\lambda$ are temperature-independent
coefficients and $\theta$ is the angle of rotation about the
[111]-direction.
We next ``freeze-in'' the NiO$_6$ rotation pattern [inset of Fig.\ 5
(b)] which corresponds to the $A_{1g}$ Raman-mode of the $R\bar{3}c$
phase.
In Fig.\ \ref{Fig.3}, we show our calculated values of the total
energy versus NiO$_6$ rotation angle $\theta$ computed at the LSDA
equilibrium volume.
Our fit of the data to Eq.\ \ref{eq:landau} yields excellent agreement and
approximately corresponds to the free energy at zero temperature
$G$($\theta,T=0$~K).
From the minimum of the free energy ($\partial G/\partial
\theta$)$_{T}=0$, we obtain the $T$ dependence of $\theta$:
\begin{equation}
\theta^{2}=\frac{\kappa}{2\lambda}(T_{C}-T) \,, \mathrm{for~}
T<T_{C}\, , \label{eq:theta_angle}
\end{equation}
and a critical rotation angle at $T=0$~K of $\theta_C=10.08^\circ$
(the optimized $\theta$ from the LSDA calculation).

We next calculate the energy stabilization obtained from freezing-in
the NiO$_6$ rotation at 0~K as the difference between cubic and
rhombohedral phases, $\Delta E =177$~meV/10-atom unit cell given by
the well depth from the total energy calculations [Fig.\
\ref{Fig.3}(a)].
With increasing temperature, the energy stabilization from the
structural distortion decreases until at $T=T_{C}$,  the thermal
excitation energy is equivalent to $\Delta E$ and the double-well
potential becomes a single well with one minimum  at
$\theta=0^\circ$.
With these conditions, we write
\begin{equation}
\Delta E=(T \cdot \Delta S)_{T=T_{C}}\, ,
\end{equation}
where $\Delta S$ refers to the entropy difference between the cubic and
rhombohedral phases.
Since LaNiO$_{3}$ has a Debye temperature $\Theta$$_{D}$ of
420~K\cite{Rajeev_et_al:1991}, at the structural transition
$T=T_{C}$, T$_{C} \gg \Theta_{D}$, we are able to confidently
treat each lattice mode as an independent harmonic oscillators which
subsequently contributes $Nk_{B}T$ to the free energy.
For the 10-atom rhombohedral LaNiO$_3$ unit cell, a single soft mode
($N=1$) describes the transition into the high-temperature cubic phase;
therefore, $\Delta S \simeq k_{B}$.
From these conditions, we calculate the LSDA critical temperature
for the structural phase transition $T_{C}$ to be  2057~K, which is
close to experimental\cite{Chaban/Kreisel_et_al:2010} result
($T_{C}$=1780~K).
The overestimation of $T_{C}$ likely originates from the calculated
enhancement of $\theta_C$, which increases the energy difference
between cubic and rhombohedral phases. %%

Using this calculated $T_{C}$ value, and combining with our earlier
calculated DFT total energy results, we obtain the coefficients in
Eq.~(6): $\kappa$=1.696~$\mu$eV/K$\cdot$$(^\circ)^2$ and
$\lambda=17.1$~$\mu$eV/$(^\circ)^4$.
These values combined with Eq.~\ref{eq:theta_angle} give the
second-order temperature dependence of the NiO$_6$ octahedral
rotational angle $\theta (T)$ shown in Fig.\ \ref{Fig.3}(b).

\subsubsection{Correlation of volume expansion with phase-transition}

With increasing temperature, the LaNiO$_{3}$ lattice undergoes a
thermal volumetric expansion process. To precisely simulate the
temperature induced rhombohedral-to-cubic phase transition process,
it is necessary to evaluate the effect of volume expansion on the
NiO$_6$ octahedral rotations across the structural phase transition.

By including the coupling between the order parameter $\theta$ and the
equilibrium atomic volume $V$, we extend the minimal 1D Landau model
discussed in the previous subsection into a 2D case, with the free
energy $G$ given as:
\begin{equation}
G(\theta, V, T)=G_{0}(V,
T)+\kappa(V)(T-T_{C})\theta^{2}+\lambda(V)\theta^{4}\, ,
\label{eq:landau_V}
\end{equation}
where $V$ is the equilibrium LaNiO$_{3}$ volume at a given temperature,
and the coefficients $\kappa$ and $\lambda$ are expressed as a function
of $V$. From 0~K to room temperature, we assume a linear volumetric thermal
expansion: $V(T)=V_{0} \times (1+\alpha_VT)$, where
$V_{0}$ is the volume at 0~K and the thermal volumetric expansion
 coefficient $\alpha_V=1.624\times10^{-5}/\textrm{K}$ is
from experiment.\cite{Chaban/Kreisel_et_al:2010}
To correlate the DFT total energy results with the Landau free energy at
0~K, we rewrite Eq.~\ref{eq:landau_V} as
\begin{equation}
G(\theta, V, T=0~K)=G_{0}(V)+A(V)\theta^{2}+B(V)\theta^{4}\, .
\end{equation}
This form indicates a practical route to calculate the optimal NiO$_6$ rotation
angle $\theta_{C}$ with respect to $V$.
We apply the procedure described in the previous subsection, but now
obtain volume-dependent coefficients $A$ and $B$ by fitting
each coefficient through a Taylor expansion about the equilibrium volume:
\begin{eqnarray}
A(V)&=&A_{0}+A_{1}(\frac{\Delta V}{V_{0}})+A_{2}(\frac{\Delta
V}{V_{0}})^{2}+\dots  \, , \\ \nonumber
B(V)&=&B_{0}+B_{1}(\frac{\Delta V}{V_{0}})+B_{2}(\frac{\Delta
V}{V_{0}})^{2}+\dots
\end{eqnarray}
where $\Delta V=V-V_{0}$.
With this method, we obtain the following coefficients:
$A_{0}=-3.6$, $A_{1}=-16.6$, $A_{2}=1600.1$~meV/$(^\circ)^2$, and
$B_{0}=17.9$, $B_{1}=-15.5$, $B_{2}=-1664.0$~$\mu$eV/$(^\circ)^4$.
This treatment of changes in the lattice volume at the
rhombohedral-to-cubic transition reveals that the magnitude of our
phenomenological coefficients $A$ ($B$) increase (decrease)
with increasing cell volume.
Thus, the curvature of the double-well potential [Fig.~\ref{Fig.3}]
becomes steeper and its depth deeper as the cell volume increases; this
fact indicates that the thermal expansion effect will hinder
the rhombohedral-to-cubic phase transition in LaNiO$_3$.
Nonetheless, the extent of the thermal
volumetric expansion effect is very limited, i.e.\ from 0~K to room
temperature, $\Delta V/V_{0}\simeq 0.5\%$, which leads to
changes in the coefficients $A$ and $B$ by less than 1\% and a deviation in our
simulated rotation angle $\theta$ at 300~K by 0.2
$^\circ$ compared to the \textit{volume-independent} model  previously described.
We conclude that, for the temperature range we investigated (0~K to
room temperature), treating the phenomenological coefficient in our
Landau theory as volume-independent suffices to produce an accurate
description of the temperature-induced octahedral phase transition
in LaNiO$_{3}$ . %%

\subsubsection{Temperature-dependent Raman Frequencies}
\begin{figure}
\centering
\includegraphics[width=0.42\textwidth]{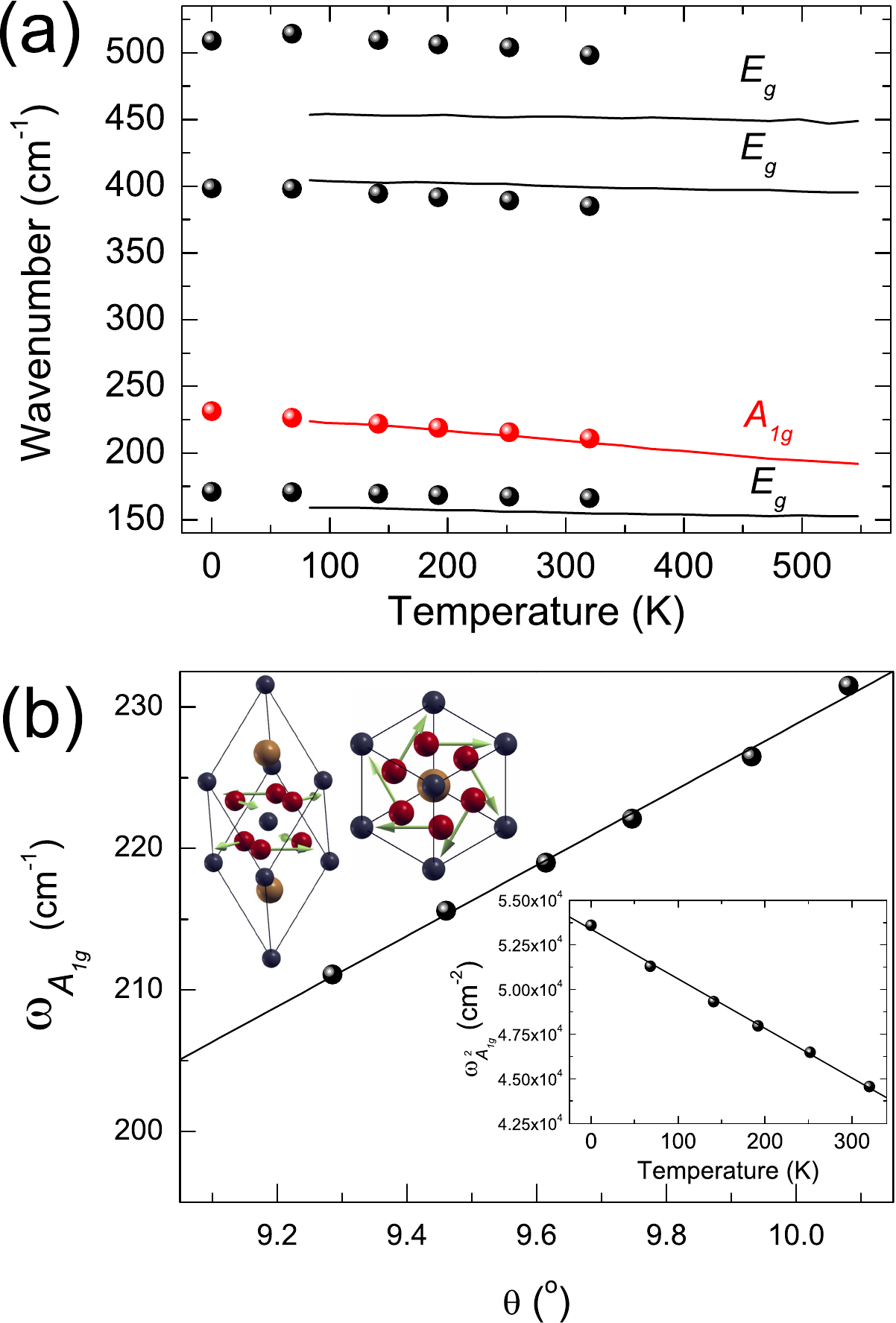}
\caption{(Color) (a) Temperature-dependent Raman frequencies for the
rhombohedral LaNiO$_3$ structures.
Solid lines indicate experimental results which are
taken from Ref.\ \onlinecite{Chaban/Kreisel_et_al:2010}.
Results from our LSDA calculations are given as the filled circles.
(b) We observe linear scaling behavior between A$_{1g}$ Raman frequency
and the NiO$_6$ octahedral rotation order parameter $\theta$.
The fitted solid line is a guide to the eye.
The inset (upper left) shows the vibrational pattern of A$_{1g}$
mode (both side and top view), while the other inset (lower right)
shows a linear change in the squared A$_{1g}$ soft-mode frequency
with temperature.} \label{Fig.4}
\end{figure}

We next study the structural and vibrational properties of
LaNiO$_{3}$ at finite temperatures. As the temperature increases
the NiO$_{6}$ rotation angle $\theta$ decreases [Fig.\ \ref{Fig.3}(b)] and
the Ni--O bond length increases.
To capture this effect in our simulation, we choose to combine the
cell parameters $a$ and $\alpha$ for LaNiO$_3$ at a given
temperature with a specific NiO$_{6}$ rotation angle $\theta$ and
Ni--O bond length [$d$(Ni--O)]. The rotation angle $\theta$ for a
given temperature is obtained from  Eq.\ \ref{eq:theta_angle}. While
for $d$(Ni--O), we refer to the experimental temperature-dependent
neutron scattering results.\cite{Lacorre/Torrance:1992} To sample
the $d$(Ni--O)--temperature space at experimental values not
available, we assume linear thermal expansion.
We then relax only the internal coordinates and use the structures that yield
the best experimental agreement to represent a snapshot of the
experimental LNO structure.
We use these structures to perform phonon calculations  and obtain
the temperature-dependent Raman frequencies [Fig.\ \ref{Fig.4}(a)].
A comparison of the experimental\cite{Chaban/Kreisel_et_al:2010} data to
our DFT calculated temperature-dependent Raman frequencies reveals good agreement:
both data sets show red-shifts in the Raman frequencies with increasing
temperature.
The calculated high-frequency $E_g$ mode, which corresponds to
stretching of the Ni--O bonds, however is systematically
underestimated by around 10\%.
The LSDA also predicts that the frequencies of the bending and
stretching $E_g$ modes decrease nearly twice as fast as that
experimentally observed. %%

We now connect the frequency of the  $A_{1g}$
Raman mode ($\omega_{A_{1g}}$) and the order parameter $\theta$
described in the previous subsections.
Here, we assign the lattice mode to the stiffness $\kappa$, or the
curvature (second-order derivative) of the potential well, as
\begin{equation}
\omega_{A_{1g}}^{2}\propto\frac{\partial^{2}G}{\partial\theta^{2}}=4\kappa(T_{C}-T)
\,, \mathrm{for~} T<T_{C}\, .
\label{eq:omega_a1g}
\end{equation}
By  comparing Eq.~\ref{eq:theta_angle} with Eq.~\ref{eq:omega_a1g}, we find
that $\omega_{A_{1g}} \propto |\theta|$ and in the temperature limit
$T \rightarrow T_{C}$ both $\theta$ and $\omega_{A_{1g}}$ approach zero.
Using our fitted value of $\kappa$, we show in Figure~\ref{Fig.4}(b)
that $\omega_{A_{1g}}$ varies linearly with $\theta$.
We also predict from the slope of Fig.~\ref{Fig.4}(b)  that the
Raman-mode shift due to changes in the NiO$_6$ rotation  angle is
$\omega_{A_{1g}}/\theta=23.0$~cm$^{-1}$/$(^\circ)$.
Finally, we plot the temperature-dependence of the squared $A_{1g}$
frequency [Fig.~\ref{Fig.4}(inset)] and find that
$\omega^{2}_{A_{1g}}$ decreases linearly with temperature,
confirming the $\omega^{2}_{A_{1g}}$($T$) relation in Eq. (12). On
the basis of these findings, we conclude that the $A_{1g}$ soft mode
is an excellent signature for the magnitude of the octahedral
rotations in rhombohedral LaNiO$_{3}$ and its deviation from cubic
phase.
%%%%%%%%%%%

%%%
\subsection{Correlation effects on the atomic structure and Raman phonon modes}
We focus in this section on how the Raman active mode frequencies
are modified through changes in electron-electron correlations.
We first decompose the effect of correlation through the Hubbard $U$
term on the structural degrees of freedom by fixing the lattice
parameter to that of the experimental  $R\bar{3}c$ structure
and allowing the internal atomic positions to fully relax.
The results of our atomic relaxations for LSDA$+U$ values of
0, 3, and 6~eV are summarized in Table \ref{tab:correlated_table}.
In all cases, the LSDA$+U$ functional accurately reproduces the
known experimental ground-state structure with the minor caveats we
discuss next. %%

With increasing correlation, we find that the NiO$_6$ rotation angle
increases beginning from the LSDA structural ground-state
($U=0$~eV), which slightly underestimates the rotation angle, to
$U=3$~eV which overestimates it by approximately 1$^\circ$.
By further increasing the Hubbard $U$ value to 6~eV, we find the
rotation angle decreases.
The consequence of keeping the unit cell volume and rhombohedral
angle fixed is that the change in the NiO$_6$ rotation angle must be
accommodated by bond stretching (or compression) rather than through rigid
rotations (constant Ni--O bond lengths).
Because the Ni atoms also occupy the $2b$ Wyckoff position with $\bar{3}$ site
symmetry, all Ni--O bond lengths are required to be equivalent.
We thus observe that our calculated Ni--O bond lengths
respond proportional to the NiO$_6$ rotation angle $\theta$
(Table \ref{tab:correlated_table}).

\begin{table}[b]
\caption{Optimized internal structural parameters of $R\bar{3}c$
LaNiO$_3$ at the experimental\cite{Lacorre/Torrance:1992} cell
parameter $a$ and rhombohedral angle $\alpha$, calculated with the
LSDA$+U$ functional and PAW pseudopotentials as implemented in {\sc
vasp}.}
\begin{ruledtabular}
\centering
\begin{tabular}{ldddd}
                    &  & \multicolumn{3}{c}{LSDA$+U$ ($U$ in eV)} \\
\cline{3-5}
       &  \multicolumn{1}{c}{Exp.} & \multicolumn{1}{c}{0} & \multicolumn{1}{c}{3}     & \multicolumn{1}{c}{6}   \\
\hline
$x$                                   & 0.797    & 0.796    & 0.800   & 0.797  \\
$d$(Ni--O) ({\AA})                    & 1.933    & 1.932    & 1.935   & 1.933  \\
$\angle$Ni--O--Ni ($^\circ$)          & 164.8    & 165.2    & 163.9   & 164.6  \\
$\theta$ NiO$_6$ rotation ($^\circ$)  & 9.20     & 8.97     & 9.77    & 9.33 \\
\end{tabular}
\end{ruledtabular}
\label{tab:correlated_table}
\end{table}

\begin{figure}[t]
\centering
\includegraphics[width=0.48\textwidth]{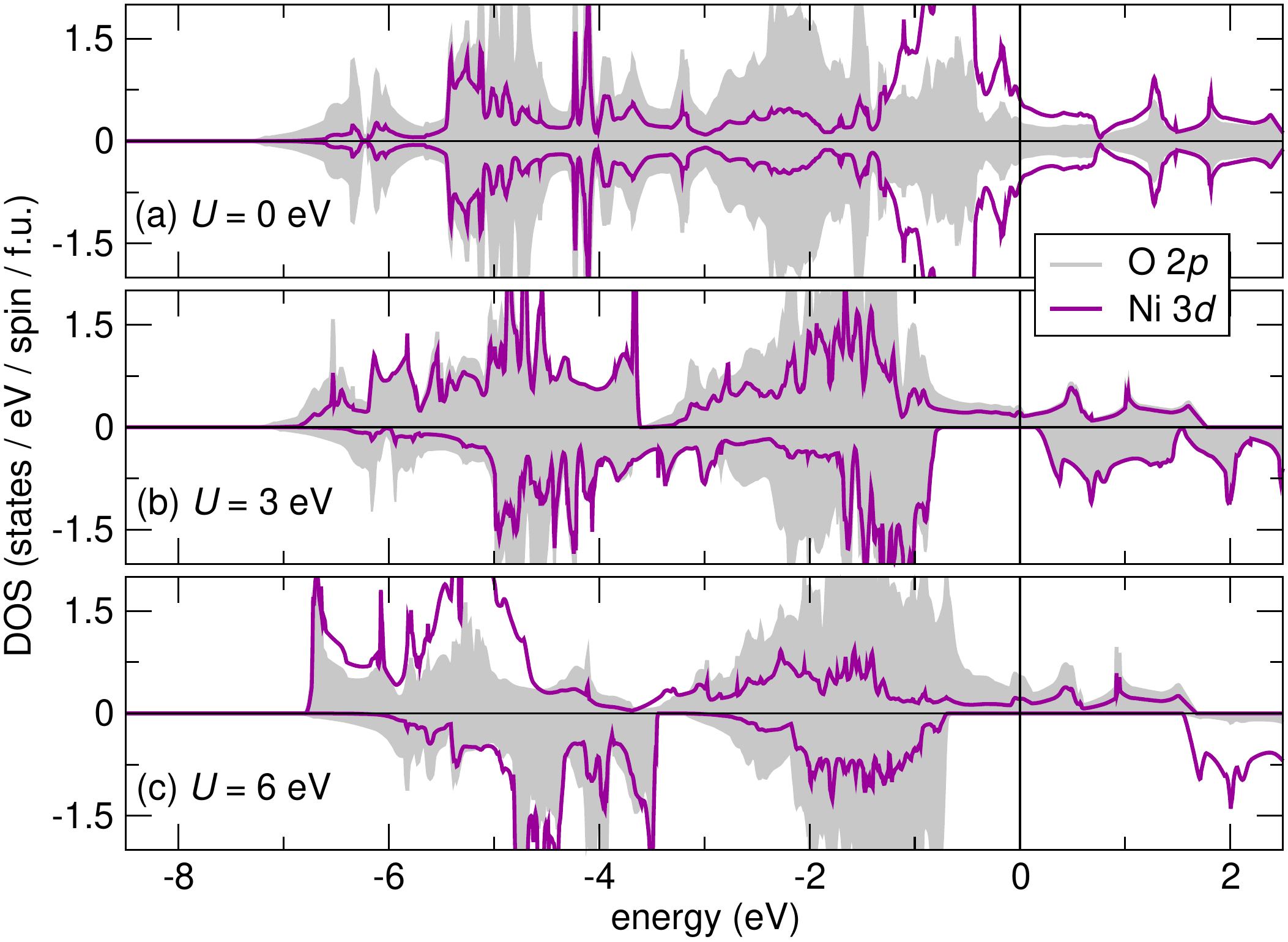}\vspace{-8pt}
\caption{(Color) Spin- and atom-resolved densities of states (DOS)
for the LaNiO$_3$ structures reported in Table
\ref{tab:correlated_table} with the LSDA$+U$ exchange-correlation
functional.} \label{fig:vasp_dos}
\end{figure}

We next examine the change in the electronic structure with
increasing correlation to evaluate how the electronic states around
the Fermi level are modified.
We show in Fig.\ \ref{fig:vasp_dos} the electronic
densities-of-states (DOS) as a function of the Hubbard $U$ value
obtained by LSDA$+U$ calculations with the {\sc vasp} code.
Consistent with earlier band structures calculations on bulk
LaNiO$_3$ with the LSDA ($U=0$~eV)
functional,\cite{Sarma/Shanthi/Mahadevan:1994,Anisimov/Bukhvalov/Rice:1999,May/Rondinelli:2010}
we find a non-magnetic ground-state with localized Ni 3$d$ states
peaked centered 1.0~eV below the Fermi level ($E_F$).
A set of delocalized Ni 3$d$ states cross $E_F$, while the O 2$p$
states are distributed throughout the entire valence band.

Here, we apply a Hubbard $U$ term on the Ni-$d$ states and
anticipate the on-site Coulomb interaction to localize the itinerant
Ni-$d$ electrons at $E_F$ and stabilize a ferromagnetic (FM) spin configuration.
Indeed, we find a half-metallic FM ground-state (1~$\mu_B$/f.u.).
Note, this configuration, however, has not been experimentally
reported.
For $U=3$~eV, we find that the DOS at the Fermi level is
substantially suppressed compared to the LSDA, with the Ni $d$
states shifting to lower energy.
Due to the increased rotational angle of the NiO$_6$, we find a
pseudo-gap opens in the majority spin DOS around -3.5~eV with
reduced hybridization between the Ni 3$d$ -- O 2$p$ states.
With a further increase of the $U$ value [Fig.\
\ref{fig:vasp_dos}(c)], we find that most of the majority spin Ni
3$d$ electrons are shifted completely to the bottom of the valence
band around -5.5~eV.
The states at $E_F$ are now mainly O 2$p$ character with a small
contribution from the Ni 3$d$ electrons. %%

\begingroup
\squeezetable
\begin{table}[t]
\caption{Raman-active vibrational modes computed using the
LSDA$+U$ formalism at the experimental cell volume are
compared to the measured values. All mode frequencies are given
in wavenumbers (cm$^{-1}$).
The experimental values are taken from
Ref.\ \onlinecite{Chaban/Kreisel_et_al:2010}.
\label{tab:fixv_raman}}
\begin{ruledtabular}
\centering
\begin{tabular}{lccccc}%
           &           $E_g$           &           $E_g$           &         $A_{1g}$              &           $E_g$           &           $E_g$       \\
\hline
Exp.       &           ---             &           156             &           209                 &           399             &           451         \\
\hline
$U=0$~eV   &           61.8            &           155.0           &           215.4               &           372.6           &           451.7       \\
$U=3$~eV   &           94.3            &           164.8           &           245.5               &           387.3           &           413.5       \\
$U=6$~eV   &           54.6            &           156.7           &           219.4               &           316.0           &           381.8       \\
\end{tabular}
\end{ruledtabular}
\end{table}
\endgroup
%%%%%%%%%%%%%%%%

\begin{table*}
\caption{%
Fully optimized structural parameters of $R\bar{3}c$ LaNiO$_3$
(rhombohedral setting), calculated from the various
exchange-correlation functionals. }
\begin{ruledtabular}
\centering
\begin{tabular}{lddddddd}%
                    &  & \multicolumn{3}{c}{VASP: LSDA$+U$ ($U$ in eV)} & \multicolumn{3}{c}{QE} \\
\cline{3-5} \cline{6-8}& \multicolumn{1}{c}{Exp.} & \multicolumn{1}{c}{$0$} & \multicolumn{1}{c}{$3$}  & \multicolumn{1}{c}{$6$} &  \multicolumn{1}{c}{LSDA}  & \multicolumn{1}{c}{GGA-PBE} & \multicolumn{1}{c}{PBEsol} \\
\hline
$x$                                 & 0.7968      & 0.787   & 0.795    & 0.792  & 0.801  & 0.813  & 0.807 \\
$a$ ({\AA})                         & 5.3837      & 5.303   & 5.308    & 5.319  & 5.324  & 5.407  & 5.367 \\
$\alpha_{\textrm{rho}}$ ($^\circ$)  & 60.8596     & 60.72   & 60.92    & 60.83  & 61.39  & 61.43  & 61.66 \\
$d$(Ni--O) ({\AA})                  & 1.933       & 1.896   & 1.905    & 1.906  & 1.923  & 1.964  & 1.952 \\
$\angle$Ni--O--Ni ($^\circ$)        & 164.8       & 167.9   & 165.6    & 165.5  & 163.3  & 159.6  & 159.6 \\
$\theta$ NiO$_6$ rotation ($^\circ$)& 9.2         & 7.35    & 8.76     &  8.19  & 10.08  & 12.31  & 11.17 \\
$\Omega$~({\AA$^3$}/f.u.)           & 56.2386     & 53.57   & 53.97    & 54.20  & 55.03  & 57.71  & 56.69 \\
\end{tabular}
\end{ruledtabular}
\label{tab:rel_structures}
\end{table*}

We now compute the Raman active lattice modes for the structures
listed in Table \ref{tab:correlated_table} to explore the change of
Raman frequencies with an increasing Hubbard $U$ value. %%
We begin by comparing the experimentally measured values to our
calculated ones (Table \ref{tab:fixv_raman}).
The best agreement with the experimental data is for the LSDA ($U=0$~eV)
exchange-correlation functional.
A Hubbard $U$ value of 3~eV overestimates the low frequency Raman
mode with $E_g$ symmetry (La anti-parallel displacements), while it
underestimates the two high frequency $E_g$ modes (Ni--O bond
bending and stretching).
%

%The lower frequency mode is related to the La cation anti-parallel
%displacements, and our calculations predict this mode is stiffer,
%probably due to the reduced covalency from the addition of the
%Hubbard $U$.
%
%In contrast, the underestimation of the two high-frequency $E_g$ modes
%(Ni--O bond bending and stretching), likely arises from a reduced
%restoring force within the LSDA$+U$ calculations as a result of reduced
%Ni states at $E_F$ with increasing $U$ (Fig.\ \ref{fig:vasp_dos}).
%

%%
Interestingly, with the LSDA$+U$=6~eV functional we find an unstable
(imaginary) zone-center phonon ($246i$~cm$^{-1}$) with $A_{2g}$
symmetry indicating that the rhombohedral structure with the simple
$a^-a^-a^-$ tilt pattern is unstable.
The atomic displacement pattern of the imaginary mode corresponds to
a three-dimensional checkerboard arrangement of dilated and
contracted octahedra---the so called octahedral ``breathing'' mode
that often accompanies charge disproportionation
reactions.\cite{Saha-Dasgupta/Popovic/Satpathy:2005}
This mode, however, has not been observed in any temperature-dependent
x-ray studies on LaNiO$_3$.
We therefore suggest that this correlation-induced octahedral
distortion is a spurious artifact of using too large of an on-site Coulomb
repulsion interaction in the LSDA$+U$ calculation.
We conclude that a Hubbard $U$ value less than 6~eV should be used
when simulating LaNiO$_3$, because of the overall poor accuracy in
the calculated Raman modes (Table \ref{tab:fixv_raman}) despite such
large Hubbard values closely reproducing the experimental lattice
parameters of LaNiO$_3$ phase (Table~\ref{tab:correlated_table}). %%

We now explore how changes in the electron correlations modify the
lattice volume by fully relaxing both the internal atom positions
and the rhombohedral structure (Table \ref{tab:rel_structures}).
For the LSDA$+U$ exchange-correlation potential, we qualitatively find the
same structural trends as in Table \ref{tab:correlated_table}.
%
%We do, however, observe an underestimation in the primitive cell volume
%when the lattice is allowed to fully relax.
%
%Due to the underestimation in the lattice volume, the calculated NiO$_6$ rotation
%angle is also substantially underestimated.
%
While the LSDA underestimates the atomic volume, when we increase the
value of $U$, the cell volume increases; this can be
understood as a result from enhanced electrostatic repulsion.
We also compute the LNO ground-state structures with the LSDA, PBE
and PBEsol functionals and norm-conserving pseudopotentials as
implemented in {\sc QE}. Typically, the LSDA underestimates the lattice
constant $a$, cell volume and Ni--O bond length, but it closely
reproduces the experimentally measured Ni-O$_{6}$ octahedral
rotation angel $\theta$. For the PBE functional, we find that both
the bond length and octahedral rotation angles are overestimated.
On the other hand, the PBEsol gradient-corrected functional corrects
some of above-mentioned overestimation from PBE, although still
slightly overestimating the rotation angle, it does provide the best
agreement with the experimental lattice constant $a$, Ni--O bond
length and equilibrium atomic volume, among all the functionals we
used for the structural optimization.

\begingroup
\squeezetable
\begin{table}[b]
\caption{Raman-active vibrational modes computed using the various
exchange-correlation functionals at the relaxed cell parameters (Table~\ref{tab:rel_structures})
compared to the experimental values.} \label{tab:relv_raman}
\begin{ruledtabular}
\centering
\begin{tabular}{lccccc}%
           &           $E_g$           &           $E_g$           &       $A_{1g}$                &           $E_g$           &           $E_g$           \\
\hline
Exp.       &           ---             &           156             &           209                 &           399             &           451             \\
\hline
$U=0$~eV   &           72.5            &           166.4           &           196.8               &           374.1           &           519.4           \\
$U=3$~eV   &           73.4            &           164.0           &           221.4               &           388.0           &           399.4           \\
$U=6$~eV   &           66.0            &           159.6           &           201.2               &           330.3           &           387.3           \\
\hline
LSDA-PZ    &           81.6            &           171.0           &           231.5               &           398.5           &           509.3           \\
GGA-PBE    &           84.3            &           168.9           &           278.9               &           351.1           &           407.1           \\
PBEsol     &           81.6            &           169.3           &           256.3               &           399.7           &           458.3           \\
\end{tabular}
\end{ruledtabular}
\end{table}
\endgroup

We now use these ground-state structures listed in Table
\ref{tab:rel_structures} and calculate their Raman-active mode
frequencies to examine the effect of lattice volume on the mode
frequencies.
Comparing these LSDA$+U$ results (Table \ref{tab:relv_raman}) to our
previously calculated lattice modes obtained using the same functionals but
at the experimental volume (Table \ref{tab:fixv_raman}), we find
that the LSDA ($U=0$~eV) functional provides the best agreement
with experimental data.
As before, we find an unstable zone-centered NiO$_6$ octahedral
breathing mode in our calculations with $U$=6 eV.
Interestingly, the LDSA$+$Hubbard $U$ method accurately predicts the
ground-state structural properties; however, it also leads to poor
predictions for the lattice normal modes. We note that in this case,
simply reproducing the correct ground-state atomic structure is not
a sufficient criterion to evaluate the performance of a functional.
This caveat is important to consider especially in the
first-principles search for perovskite materials with large
electron--phonon interactions. %%

We also find a close relation between the predicted Raman phonon
frequencies and structural parameters of the bulk rhombohedral LNO,
especially for the $A_{1g}$ and the two high-frequency $E_g$ modes.
Typically, the $A_{1g}$ mode frequency is sensitive to the
octahedral rotation angle $\theta$, whereas the bond bending and
breathing $E_g$ modes are substantially more sensitive to the
predicted Ni--O bond length and cell volume $\Omega$, respectively.
The LSDA underestimates both the lattice constant $a$ and the cell volume
$\Omega$, but it also overestimates the rhombohedral angle
$\alpha_{\textrm{rho}}$. This leads to a cancellation in
errors and an overall good prediction of both $\theta$ and $d$(Ni--O) and
therefore calculated rotation ($A_{1g}$) and bending
($E_g$)  phonon frequencies that agree very well with the experiment.
The LSDA functional only overestimates the stretching $E_g$-mode
frequency (by 10\%) due to the underestimation of cell volume.
Compared to the LSDA, the PBEsol functional
improves prediction of the structural parameters, such as $a$ and
$\Omega$ (Table~\ref{tab:rel_structures}).
Except for the overestimation of $\theta$, PBEsol also
predicts a $d$(Ni--O) bond lengths close to experiment.
Therefore, as shown in Table~\ref{tab:relv_raman}, the
two high-frequency $E_g$ modes are in good agreement with
experiment, but the $A_{1g}$ mode is overestimated by more than 20\%.
The PBE functional, in contrast,
overestimates all of the major bulk LNO structural parameters, thus providing
the poorest description of the Raman frequencies
among the three functionals explored in the QE code.
To summarize, both the LSDA and PBEsol functionals predict overall
adequate Raman frequencies for LNO that are close to experiment;
however, each functional still has deficiencies for specific
vibrational modes, and in fact the LSDA values are closest to
experiment overall.
%%%

%
We emphasize here that the NiO$_6$ rotation angle $\theta$, the
order parameter to characterize the structural phase transition in
LaNiO$_3$, is highly sensitive to the exchange-correlation
functional.
Even calculations with same functional, but different
pseudopotentials (for example the  LSDA calculations performed with
the {\sc vasp} and {\sc QE} codes) yield $\theta$ values with
obvious differences.
Therefore we suggests that an accurate and comprehensive study of
various theoretical approximations on the descriptions of the octahedra
rotation angles in rhombohedral perovskite oxides is needed.

%%%%
\subsection{Electronic structure and experimental energy spectra}

\begin{figure}
\centering
\includegraphics[width=0.30\textwidth]{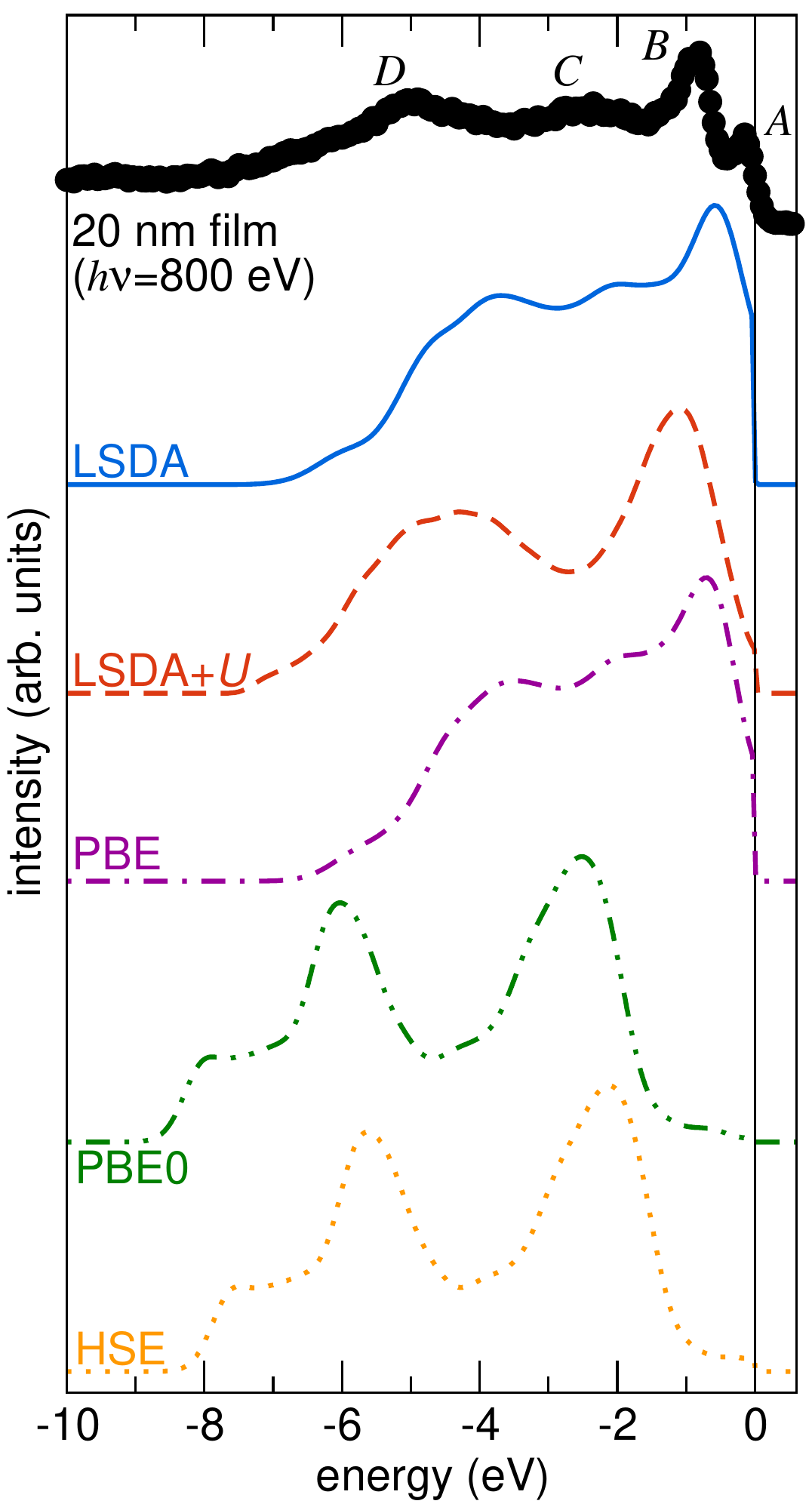}
\caption{(Color) The experimental PES spectrum for a 20~nm
crystalline LNO film (filled circles)  from Ref.\
\onlinecite{Horiba_permission:2010} is compared to the density of
states calculated with the LSDA+$U$ (U=5.74 ~eV), gradient-corrected
(PBE) and hybrid (PBE0 and HSE) exchange-correlation functionals.
The calculated data are smeared with Lorentzian and Gaussian functions and
truncated with a Fermi-Dirac distribution to facilitate the comparison.
See the main text for peak position assignments.} \label{fig:exp_compare}
\end{figure}

In this section we compare our first-principles results with recent
photoemission spectroscopy (PES) data to identify the degree of
electronic correlations in rhombohedral LaNiO$_3$.
In a single electron picture, PES measures the excitation energies
for non-interacting electrons from the valence band into the
continuum and therefore can be used as a first-order reference to
single-particle DFT studies. We show in Figure \ref{fig:exp_compare}
experimental PES data\cite{Horiba_et_al:2007} from a crystalline
20~nm LaNiO$_3$ film, and compare it with our first-principles
calculated valence band DOS to evaluate the accuracy of our
calculations in reproducing the known electronic structure.
In order to make a more accurate comparison, we first smear our
calculated DOS with a Gaussian function (FWHM$=$0.20~eV) to account
for the experimental resolution and multiply by a 20~K Fermi-Dirac
distribution; then we convolute an energy-dependent Lorentzian
function [full width at half-max (FWHM)$=0.1|E-E_F|$~eV,
\cite{Sakurai:Book} where $E_F$ is the Fermi level] with our
calculated DOS to include lifetime broadening effects of the
photon-excited electrons.

The experimental PES consists of four main features: peaks {\sc a}
and {\sc b} are assigned to the Ni $e_{g}$ and $t_{2g}$ states, and
the deeper {\sc c} and {\sc d} spectra to the O 2$p$ dominant
states.\cite{Horiba_et_al:2007}
The $e_{g}$ states ({\sc a}) are clearly resolved as a sharp peak at
the Fermi level and the strong $t_{2g}$ peak is located at 1.0~eV
below the Fermi edge. The O 2$p$ states ({\sc c}), located below the
$t_{2g}$ states, correspond to non-bonding O 2$p$ states, as their
hybridization with the Ni 3$d$ states is restricted by symmetry.
The O 2$p$ bonding states ({\sc d}) are much broader and located
between -8 and -4~eV. Compared to experimental PES results, none of
the exchange correlation functionals are able to reproduce the sharp
spectral intensity of the $e_g$ peak at the Fermi level.
Each method leads to an over delocalization of the itinerant $e_g$
electrons (Fig.\ \ref{fig:exp_compare}).

\begin{figure}
\centering
\includegraphics[width=0.48\textwidth]{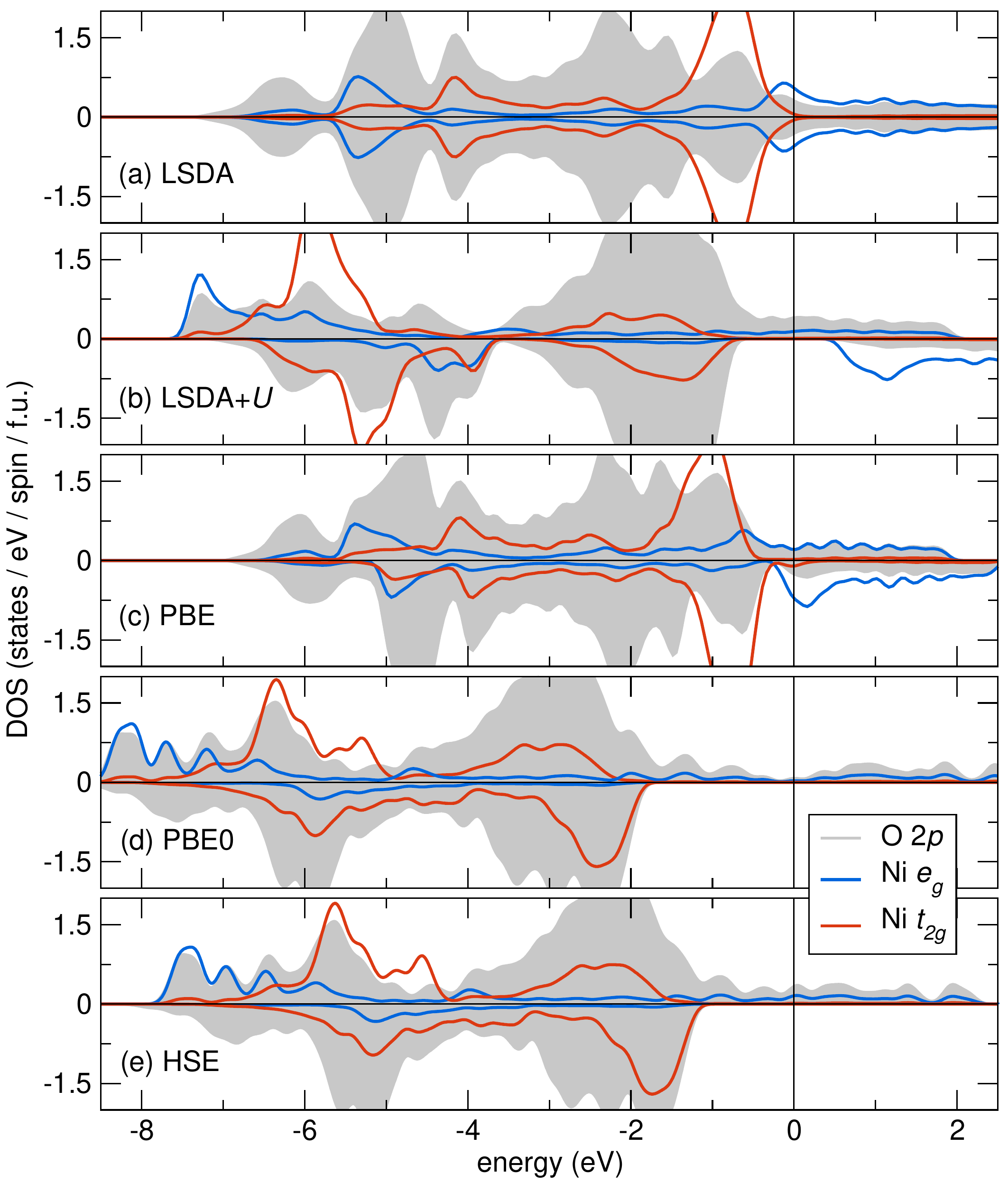}
\caption{Spin- and orbital-resolved density of states for
rhombohedral LaNiO$_3$, obtained from LSDA, LSDA+$U$ ($U$=5.74~eV),
PBE, PBE0 and HSE  calculations. The Fermi energy is the energy
zero. In each case the structure is fully relaxed according to the
specific functional with the exception that the DOS for the hybrid
XC-functionals are calculated with the LSDA ground-state atomic
structure.} \label{fig:dos_all_composite}
\end{figure}

Despite the inability to reproduce peak {\sc a}, the LSDA functional
does exceedingly well in reproducing the experimental valence band
features---both the correct valence bandwidth and energy peak
positions. As shown in Fig.\ \ref{fig:dos_all_composite}(a), the
LSDA predicts a pronounced hybridization between the Ni $t_{2g}$ and O
2$p$ states almost throughout the full valence band. The $e_g$
states, located at the Fermi level, are separated in
energy from the Ni $t_{2g}$ orbitals.

The LSDA$+U$ functional is also able to reproduce the correct
bandwidth; however, it fails to predict the correct energy peak
positions. It shifts the $t_{2g}$ states (peak {\sc b})
toward the bonding O 2$p$ states (peak {\sc d}) as observed in the
orbital resolved DOS [Fig.\ \ref{fig:dos_all_composite}(b)].
There is an additional shift of bonding O 2$p$ spectral weight from the bottom
of the valence band to around feature {\sc b} which gives the impression that
the localized $t_{2g}$ states below the Fermi level are only broadened and
not shifted.
These two effects combine to give a reduced intensity in the non-bonding 2$p$
states (peak {\sc c}), where as experimentally they contribute greater intensity
to the PES data.
The PBE functional also predicts the correct delocalized electronic
characters for LaNiO$_{3}$, but underestimates the valence bandwidth
[Fig.\ \ref{fig:dos_all_composite}(c)]. Moreover, it predicts a
ferromagnetic ground-state (0.56 $\mu_{B}$ local magnetic moment per
Ni), in contrast to the paramagnetic LSDA results. Therefore, the
PBE functional predicts an electronic structure that is intermediate
between the LSDA and LSDA$+U$ methods. %%

Both hybrid functionals give poor agreement between the calculated
DOS and the experimental PES data.
In each case the bandwidth is overestimated, with the PBE0 (HSE)
functional $\approx$ 1.5~eV (1~eV) larger than the experimental
results. This is in contrast to the PBE functional which produces a
narrow ($\sim$6~eV wide) valence band [Fig.\
\ref{fig:dos_all_composite}(c)].
As mentioned earlier, similar errors are also found in other
itinerant magnetic metals when using the PBE0 and HSE functionals.\cite{Paier:2006}
The major peaks are also red-shifted by approximately 2~eV, i.e.
PBE0 and HSE functionals shift the major Ni $t_{2g}$ states to between
-7.5 and -5~eV below $E_{F}$. Similar to the LSDA$+U$ approach, the hybrid
functionals suppress the contribution of Ni states at $E_F$.
This results in the removal of the Ni $e_g$ states from the Fermi
level and shifts them to the bottom of the valence band.
This shift of electronic states produces the unusual shoulder at
-8~eV in the DOS calculated with the hybrid functionals [Fig.\
\ref{fig:exp_compare}(d) and (e)]. %%

\begin{figure}
\centering
\includegraphics[width=0.5\textwidth]{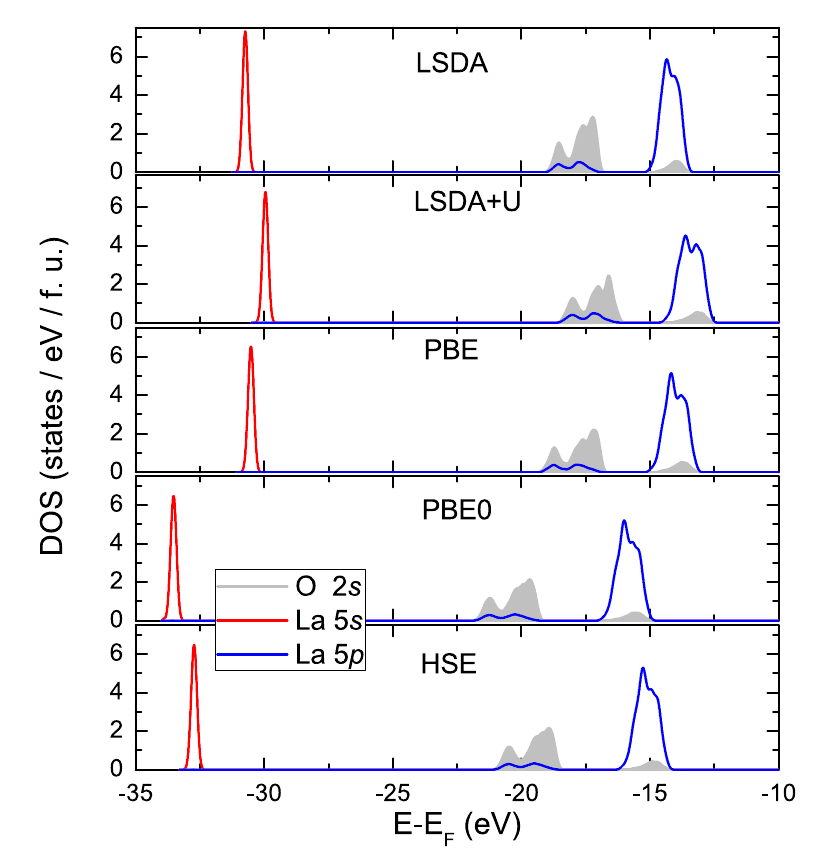}
\caption{Orbital-resolved density of states in the core energy
region obtained from LSDA, LSDA$+U$ ($U=$5.74~eV), PBE, PBE0 and HSE
calculations. The Fermi energy level is shifted to the energy zero.
Note, the binding energies predicted by PBE0 and HSE are not
obtained by a simple rigid shift of the PBE results.} \label{core_dos}
\end{figure}

The above comparison to the experimental data clearly indicates that
the beyond-LSDA methods (the LSDA$+U$ and hybrid density
functionals) incorrectly describe the electronic structure of LaNiO$_3$.
The origin for these discrepancies lies in the fact that the valence
band is primarily composed of strongly delocalized Ni $t_{2g}$ and O
2$p$ states. Therefore, metallic LaNiO$_3$ is able to strongly
screen the electron--electron interactions. This screened
electron-electron interaction can lead to renormalization of the
electronic states near the Fermi level, which are responsible for
the sharp $e_{g}$ peak observed experimentally.
\cite{Eguchi/Shin_et_al:2009} The accurate treatment of this
phenomenon in LaNiO$_3$  requires a calculation using many-body
DFT methods, i.e.\ quasi-particle \emph{GW}, which includes the
dynamically screened Coulomb interaction and therefore treats the
electron screening effect in a dynamic (energy-dependent) way.
\cite{Eguchi/Shin_et_al:2009} Our beyond-LSDA methods, such as
LSDA$+U$ and hybrid functionals,\cite{Nohara:2009,Paier:2006}
however, are only included in the exchange-correlation functional, which is
independent of the quasi-particle frequency.
As such, neither the on-site Coulomb repulsion interaction from
LDA+$U$, nor the exact-exchange energy from hybrid density functionals,
is able to reproduce the screening effects (and the above-mentioned discrepancies)
present in LaNiO$_3$.
The inability of these beyond-LSDA methods to capture the dynamic
screening effects  manifests as shifts in core-level binding
energies.
To explore this effect, we show in Fig.\ \ref{core_dos} the
calculated orbital-resolved DOS in the core energy region (-35
through -10~eV) for LaNiO$_{3}$.
The discrete core level states are primarily La 5$s$, 5$p$ and O
2$s$. They are weakly overlapped and easily distinguishable by
spectroscopic measurements.
In Table \ref{tab:xps}, we compare our calculated core-level binding energies
to recent XPS data.\cite{Masys:2010}
Although the LSDA method accurately reproduces the delocalized states
in the valence region, it largely underestimates the binding energies
of the core level states.
In contrast, both hybrid XC-functionals
significantly improve the description of the peak positions in the
core region.
Unlike the delocalized valence electrons, the core level states are
weakly modulated by the screening effect from the valence region,
and therefore exhibit  strongly localized electronic character. In
this case, the hybrid functionals, through the addition of
Fock-exchange, greatly reduce the self-interaction errors present in
the LSDA and GGA functionals and therefore substantially improve the
calculated energy spectra.
We also note that within the LSDA$+U$ framework, the Hubbard $U$
does not affect the core level states since it is applied only to
the valence Ni 3$d$ orbitals, and therefore the core level states
are only rigidly shifted in energy respect to those calculated with
the LSDA functional. %%

\begingroup
\squeezetable
\begin{table}
\caption{Calculated binding energies (in eV) for the major energy
peaks in the core region from this study are compared to those calculated
with the PBE0 (designated by the *) functional and experimental (Exp.) values, taken
from Ref.\ \onlinecite{Masys:2010}. The assignment of the peak
positions follows that of Ref.\ \onlinecite{Masys:2010}.}
\begin{ruledtabular}
\centering
\begin{tabular}{lccccccc}%
Peak   & Exp.  &  LSDA  &   PBE    & LSDA+$U$ &  HSE  &  PBE0 & PBE0$^*$  \\
\hline
b$/$c  & 16.7  &  14.2  &  14.4    &  13.6    &  15.3 &  16.0 &   17.1    \\
d      & 21.0  &  17.6  &  17.6    &  17.0    &  19.3 &  19.8 &   20.7    \\
e      & 23.6  &  18.7  &  18.6    &  18.0    &  20.4 &  21.1 &   21.8    \\
f      & 33.5  &  30.5  &  30.7    &  30.0    &  32.7 &  33.5 &   34.8    \\
\end{tabular}
\end{ruledtabular}
\label{tab:xps}
\end{table}
\endgroup

Based on our first-principles calculations, we find that strong
hybridization between the Ni 3$d$ and O 2$p$ states reduce the
on-site $d$-orbital Coulomb interaction in LaNiO$_{3}$ through
enhanced screening.
The conventional LSDA method accurately reproduces the valence band
structure and also provides the decent experimental agreement to the
structural properties. In contrast, the hybrid exchange-correlation
functionals work best for the core level states.
We do find one subtle caveat: the Ni $e_g$ states experimentally
exhibit dynamical correlation effects, which we are unable to
capture in either the LSDA, LSDA+$U$, or hybrid functionals. This
enhanced spectral weight at the Fermi level has recently been
reproduced experimentally and through LDA$+$dynamical mean-field
theory (DMFT) calculations
\cite{Stewart/Haule/Chakhalian_etal:2011,Hansmann10p235123,Georges:2011}, where a
dynamic (frequency-dependent) treatment of the Ni 3$d$
electron--electron interactions leads to an enhanced effective mass
and optical conductivity.
%%%%%

To answer the principal question posed earlier regarding the
metallic behavior in LaNiO$_{3}$:  We find that the
three-dimensional topology of the perovskite structure with
corner-connected NiO$_6$ octahedra supports strong
$pd$-hybridization and stabilizes the metallic state. Moreover, the
relatively small octahedral rotations with rhombohedral symmetry and
the weak electron-electron correlation effects are insufficient to
disrupt the metallic state.

%%%%
\section{Summary and Conclusions}
In summary, we have studied the lattice normal modes and electronic
properties of the correlated metal LaNiO$_{3}$ with first-principles
calculations using a variety of exchange-correlation functionals.
We examined the rhombohedral-to-cubic structural transition in
LaNiO$_{3}$ within second-order Landau phase transition theory.
We found that the $A_{1g}$ Raman-active mode acts as a clear descriptor for
the magnitude of the octahedral rotations in rhombohedral perovskites with
the $a^-a^-a^-$ tilt pattern.
We therefore suggest Raman spectroscopy is a plausible route for
quantifying the magnitude of octahedral rotations in rhombohedral
perovskite oxides. %%

Using a linear response method we showed that
the correlation effects in LaNiO$_3$ originate from the Ni-$e_{g}$ orbitals.
We then proposed an orbital-dependent effective Hubbard $U$ value of 5.74~eV
for LaNiO$_{3}$ to be used in the LSDA$+U$ formalism.
By comparing the results obtained from the various functionals with
experimental spectroscopic data, we found an accurate treatment of
the correlation effects in LaNiO$_{3}$ cannot be simply obtained by
the LSDA+$U$ or hybrid functional methods.
We identified that there are strong hybridization effects between the
Ni $t_{2g}$ and O 2$p$ within LaNiO$_{3}$ that result in enhanced screening
capabilities and act to reduce the electronic correlations in the e$_{g}$ orbitals.
Among the various DFT exchange-correlation functionals examined, we
find that only the LSDA is capable of reproducing \textit{both} the delocalized valence
states and the experimentally measured lattice dynamics. %%

\begin{acknowledgments}
This work was supported by the U.S.\ DOE, Office of Basic Energy Sciences,
under Contract No.\ DE-FG02-07ER46431 (GYG) and DE-AC02-06CH11357 (JMR).
IG and AMR were supported by the Office of Naval Research under
grant No.\  N00014-11-1-0578.

Computational support was provided by a DURIP grant, a Challenge
Grant from the HPCMO, and the high-performance computing facilities
at the Argonne Center for Nanoscale Materials. This work was also
partially supported by the IMI and AQUIFER Programs of the NSF under
Award No. DMR-0843934, managed by the International Center for
Materials Research, University of California, Santa Barbara, USA.
The authors thank S. J. May, J. E. Spanier, and A. Stroppa for
useful discussions.
\end{acknowledgments}

\bibliography{rondo}

\end{document}